\documentclass{aa}
\usepackage{graphics}

\newcommand{\HI}{\protect\normalsize H\thinspace\protect\footnotesize
I\protect\normalsize} 
\newcommand{\HIb}{\protect\normalsize\bf H\thinspace\protect\footnotesize\bf
I\protect\normalsize\bf} 
\newcommand{\HIc}{\protect\normalsize\it H\thinspace\protect\footnotesize\it
I\protect\normalsize\it} 
\newcommand{\HIcap}{\protect\small H\thinspace\protect\scriptsize
I\protect\small} 
 
\newcommand{\B}{{$B$}}
\newcommand{\ml}{{$M_{HI}/L_B$}}
\newcommand{\tfr}{Tully\,--\,Fisher relation}
\newcommand{\tfd}{Tully\,--\,Fisher distance}
\newcommand{\kms}{\,km\,s$^{-1}$}
\newcommand{\etal}{et~al.\ }
\newcommand{\cf}{{\it cf.}\ }
\newcommand{\eg}{{\it e.g.},\ }         
\newcommand{\ie}{{\it i.e.},\ }         

\begin{document}

   \thesaurus{03    
              (11.03.1;   
               11.03.4 Fornax; 
	       11.06.2;  
               11.07.1;  
               13.19.1)  
             }
   \title{The Neutral Hydrogen Content of Fornax Cluster Galaxies }

   \author{Anja Schr\"oder \inst{1} 
   \and    Michael J. Drinkwater \inst{2}
   \and    O.-G.~Richter \inst{3}
          }
 
   \offprints{A. Schr\"oder}
 
   \institute{Observatoire de la C\^ote d'Azur,
              B.P. 4229,
              06304 Nice Cedex 4,
              France
   \and 
             School of Physics,
             University of Melbourne,
             Victoria 3010,
             Australia
   \and 
            Hamburger Sternwarte,
            Gojenbergsweg 112, 
            20119 Hamburg,  
            Germany
                }
 
   \date{Received date; accepted date}
 
   \maketitle
 

   \begin{abstract}

We present a new set of deep \HI\ observations of member galaxies of
the Fornax cluster. We detected 35 cluster galaxies in \HI. The
resulting sample, the most comprehensive to date, is used to
investigate the distribution of neutral hydrogen in the cluster
galaxies.

We compare the \HI\ content of the detected cluster galaxies with that
of field galaxies by measuring \HI\ mass-to-light ratios and the \HI\
deficiency parameter of Solanes et al.\ (\cite{Sol96}).  The mean \HI\
mass-to-light ratio of the cluster galaxies is $0.68\pm 0.15$,
significantly lower than for a sample of \HI -selected field galaxies
($1.15\pm 0.10$), although not as low as in the Virgo cluster
($0.45\pm 0.03$).  In addition, the \HI\ content of two cluster
galaxies (NGC\,1316C and NGC\,1326B) appears to have been affected by
interactions. The mean \HI\ deficiency for the cluster is
$0.38\pm0.09$ (for galaxy types $T\!=\!1$\,--\,6), significantly
greater than for the field sample ($0.05\pm0.03$). Both these tests
show that Fornax cluster galaxies are \HI -deficient compared to field
galaxies. The kinematics of the cluster galaxies suggests that the
\HI\ deficiency may be caused by ram-pressure stripping of galaxies on
orbits that pass close to the cluster core.

We also derive the most complete \B -band \tfr\ of inclined spiral
galaxies in Fornax. A subcluster in the South-West of the main cluster
contributes considerably to the scatter. The scatter for galaxies in
the main cluster alone is 0.50\,mag, which is slightly larger than the
intrinsic scatter of 0.4\,mag. We use the \tfr\ to derive a distance
modulus of Fornax relative to the Virgo cluster of $-0.38\pm
0.14$\,mag. The galaxies in the subcluster are ($1.0\pm0.5$)\,mag
brighter than the galaxies of the main cluster, indicating that they
are situated in the foreground. With their mean velocity 95\kms\
higher than that of the main cluster we conclude that the subcluster
is falling in to the main Fornax cluster.

   \end{abstract}
 
\section{Introduction}

The nearby Fornax cluster has a high central surface density of
galaxies (Ferguson \cite{F89b}) although it is not a rich cluster in
Abell's (\cite{Abe}) sense. It forms an intermediate sample suitable
for comparison with richer clusters like Virgo (Huchtmeier \& Richter
\cite{Virgo2}; Cayatte \etal \cite{C94}) and Hydra~I (McMahon \etal
\cite{McM:etal}), as well as smaller groups and nearby field galaxies
(Huchtmeier \& Richter \cite{KKT3}; Marquarding \cite{MM00}).

\begin{table*}[bt]
\caption[]{Observational parameters of the four runs } \label{obspar}
\begin{center}
{
\begin{tabular}{lcccc}
\noalign{\smallskip}
\hline
\noalign{\smallskip}
run & \multicolumn{1}{c}{May/Jun 91} & \multicolumn{1}{c}{Jan 92} & \multicolumn{1}{c}{Jun 93} & \multicolumn{1}{c}{Apr/May 94} \\
\noalign{\smallskip}
\hline
\noalign{\smallskip}
system noise temperature & 40-45\,K & 40-45\,K & 30\,K & 30\,K \\
channel & 1024 & 2048 & 2048 & 2048 \\
bandwidth & 10\,MHz & 8\,MHz & 32\,MHz & 32\,MHz \\
channel spacing & 4.1\kms & 1.7\kms & 6.6\kms & 13.2\kms \\
velocity resolution (after smoothing) & 9.8\kms & 7.9\kms & 15.8\kms & 15.8\kms \\
\noalign{\smallskip}
\hline
\noalign{\smallskip}
\end{tabular}
}
\end{center}
\end{table*}

In particular, Fornax lies almost opposite in the sky from the
well-studied Virgo cluster but at a distance just about equal to it
(\eg Pierce \cite{P89}; Bridges \etal \cite{B91}; Hamuy \etal
\cite{H91}; McMillan \etal \cite{M93}; see also Table 6.1 in
Schr\"oder \cite{AS}).  A better understanding of the relative
distance and the substructuring in both clusters would allow one to
disentangle the effects of virgocentric infall from larger-scale
motions relative to either one or more distant galaxy aggregates, such
as the ``great attractor'' (Kolatt \etal \cite{GA}), or the microwave
background radiation reference frame. Fornax has also been used in
studies of the extragalactic distance scale (see Freedman \etal
\cite{F01}, and below).

The \HI\ content of the Fornax cluster galaxies has, heretofore -- by
virtue of the cluster's southern declination just beyond the reach of
most major northern radio telescopes -- not been studied in a
comprehensive way (\cf Sec.~\ref{data} for references).  More recent
\HI\ surveys of the cluster galaxies include Horellou \etal
(\cite{H95}) and Bureau \etal (\cite{BMS}). Barnes \etal (\cite{B97})
conducted a shallow blind survey of the inner $8\degr \times 8\degr$
of the cluster; a more detailed blind survey of the whole cluster area
using the Parkes Multibeam receiver (Staveley-Smith \etal \cite{SS96})
is currently under way (Waugh \etal \cite{W00}).

Our main motivation in this study was to obtain an improved \tfr\ for
the Fornax cluster (\cf Schr\"oder \cite{AS}) as well as confirming
cluster membership with new \HI\ radial velocities.  Our sample
included all galaxies listed in the RSA\footnote{A {\bf R}evised {\bf
S}hapley-{\bf A}mes Catalog of Bright Galaxies (Sandage \& Tammann
\cite{RSA}).} catalogue that lie within 5 degrees from the approximate
cluster centre at $\alpha = 03^h 35^m$ and $\delta = -35.7^\circ$
(basically the position of NGC\,1399) and which have a radial velocity
less than 2300\kms . To this we added those galaxies from the
comprehensive catalogue of Fornax cluster galaxies by Ferguson
(\cite{HCF}, hereafter FCC) which he (a) judged to be either certain
or possible members, and (b) classified to be of sufficiently late
morphological type to suspect a detectable \HI\ content.  We also
decided to reobserve some of the weaker and less well-determined \HI\
lines already known from the \HI\ catalogue by Huchtmeier \& Richter
(\cite{HIcat}, henceforth HR89).  Finally, two possible background
galaxies were added to confirm non-membership.

The final sample comprises 66 galaxies of which we detected 37 in \HI . 
The results are presented in Section~\ref{data} with the description
of the observations given in Section~\ref{obs}. In Section~\ref{hi} we
analyse the distribution of \HI\ content in the cluster galaxies, and
in Section~\ref{tf} we derive the \tfr .

\section{Observations} \label{obs}

The Parkes 210\,ft (64\,m) radio telescope was used during several
sessions to observe the sample of Fornax galaxies described above.
Additionally, several galaxies previously observed in the 21\,cm line
(\cf the HR89 catalogue) were reobserved to improve data quality and
to allow a better comparison with previously published data.  The
first observing run was in late May and early June 1991, a second in
January 1992, a third one in June 1993, and a fourth one in late April
and early May 1994. Table~\ref{obspar} gives an overview of the
observational parameters.

All observations were carried out in the total power mode. 10-minute
ON-source observations were preceded by an equal length OFF-source
observation at the same declination but 10.5 minutes earlier in right
ascension, so as to traverse the same path in geocentric coordinates
during both the reference and the signal observation. To avoid having
to use too many intra-cluster OFF-source positions, we attempted to
use a single OFF-source observation outside the cluster area for
several cluster galaxies loosely aligned along a ``path'' at roughly
equal declination and -- again -- spaced about 10 minutes in right
ascension.  It should, however, be noted that a number of OFF-source
positions were located well within the boundaries of the Fornax
cluster.

For each galaxy such ON/OFF-observations have been repeated until the
signal was unambiguous (in case of interferences) and the
signal-to-noise was at least 3\,--\,5.

The dual-channel AT 21\,cm receiver used in 1991 and 1992 had a system
noise temperature of order 40\,--\,45\,K.  In 1993 and 1994 a new
receiver with a system noise temperature of about 30\,K was used.  In
1991 a 1024-channel autocorrelator served as the backend.  It was
split into two banks of 512 channels each which detected the two
independent polarisations.  A bandwidth of 10\,MHz ($\rm\approx
2100$\kms ) with a central frequency corresponding to a heliocentric
radial velocity of 1300\kms\ was used for all galaxies regardless of
the availability of a known radial velocity. Beginning in 1992 a new
autocorrelator (technically quite similar to the correlators in use at
the Australia Telescope Compact Array in Narrabri) was available with
2048 channels, which were also split into 2 banks of 1024 channels and
8\,MHz bandwidth each. Once, in January 1992, this new autocorrelator
suffered a processor failure and data for a single 24-hour period were
again taken with the old autocorrelator.

The autocorrelator setup was changed in 1993 to accommodate other
parallel programs to be reported elsewhere.  The bandwidth was changed
to 32\,MHz with 1024 channels in each bank.  In 1994 we used a split
into 4 banks of 512 channels each with an IF offset of 25\,MHz for the
third and fourth bank of the autocorrelator to also cover a higher
velocity range out to 12000 $\rm km\, s^{-1}$.  This information came
basically free of cost, since banks 1 and 3 as well as 2 and 4 were
fed by the same polarisation output from the receiver, \ie would have
differed only by the quantization noise which was practically
undetectable.  No serendipitous signals, however, showed up in this
higher velocity range for Fornax galaxies.

In all cases the two different polarisations were averaged during data
reduction.  Fitted spectral baselines consisting of a polynomial of
moderate order added to a sine function with a period equal to that of
the standing wave pattern of the Parkes telescope (5.8\,MHz) were
subtracted to form the final spectra.

The online control program automatically corrected for the zenith
angle dependence of the telescope sensitivity.  The primary flux
calibration was obtained by measuring standard sources from the Parkes
catalogue (Wright \& Otrupcek \cite{Wri:etal}) at the beginning of
each observing session; it was stable to within $\approx$ 10\%.  As an
added check of system performance secondary \HI\ flux calibrators
chosen from the new compilation of \HI\ data for RSA galaxies by
Richter \etal (1994, priv.\ comm.) were observed from time to
time. Based on those data the internal consistency of the flux scale
is indeed judged to be better than 15\%.

\section{Results} \label{data}

We present our results in this section as a table of the \HI\
measurements, notes on individual galaxies and a comparison with other
measurements.  The \HI\ spectra of the detected galaxies are shown in
the Appendix.  It should be noted that the number of Hanning smoothing
operations for the displayed spectra varies from 1 to 4 depending on
the overall signal-to-noise ratio.

\subsection{The new \HIc\ Data}

In Table~2 we present the averaged data of the four runs (in general a
weighted mean) together with some parameters from the FCC. The Table
columns are as follows.

{\it Col.\ 1a:} Galaxy identification according to the FCC.

{\it Col.\ 1b:} Other galaxy identification: NGC, IC, ESO in this
order of preference.

{\it Col.\ 2:} Coordinates in J2000 equinox.

{\it Col.\ 3:} Morphological type from the FCC. Where the galaxy was
not in FCC, RC3\footnote{Third Reference Catalogue of Bright Galaxies
(deVaucouleurs \etal \cite{RC3}).} values have been taken.

{\it Col.\ 4:} \B -band magnitude from the FCC. Where the galaxy was
not in FCC, RC3 values have been taken.

{\it Col.\ 5:} Decimal logarithm of the diameter, $\log D$, in units
of $0\farcm1$ at the 26.5th \B -band isophote from the FCC. For
non-FCC members we have taken $\log D_{25}$ from the RC3 and added
0.05, which is the mean difference between FCC and RC3 values.

{\it Col.\ 6:} Axial ratios, $\log R$, from Lauberts \& Valentijn
(\cite{LV}). For galaxies with no entry we have taken the axial ratios
from Paturel \etal (\cite{P00}; \cf the LEDA\footnote{Lyon-Meudon
Extragalactic DAtabase; http://leda.univ-lyon1.fr.} database). They
are indicated with a superscript {\it a}. For one galaxy (superscript
{\it b}) the axial ratio comes from MacGillivray \etal (\cite{MG88}),
and for two galaxies no axial ratios could be found in the literature
and our own measurements are given (superscript {\it c}).

{\it Col.\ 7a:} Weighted mean heliocentric \HI\ radial velocity in
\kms\ taken at the midpoint of the \HI\ profile at the 20\% level. The
optical convention $v=c (\lambda-\lambda_o)/\lambda_o$ is used.

{\it Col.\ 7b:} Error of the radial velocity, calculated according to
Fouqu\'e \etal (\cite{F90}): $\sigma(v)=4R^{0.5}P^{0.5}s^{-1}$ where
$R$ is the resolution in \kms\ (1.2 times the channel spacing), $P$ is
the steepness of the edges of the profile ($P=(\Delta v_{20}-\Delta
v_{50})/2$), $s$ is the signal-to-noise, \ie peak $S_h$ over rms noise
level.

{\it Col.\ 8a:} Velocity width in \kms\ of the \HI\ profile measured
at the $50\%$ level of the peak intensity, $\rm\Delta v_{50}$,
corrected for instrumental broadening: no correction for $0<R \leq
8$\,\kms , $-3$\,\kms\ for $8<R \leq 16$\,\kms , $-5$\,\kms\ for $16<R
\leq 32$\,\kms , $-7$\,\kms\ for \mbox{$32<R$}. The smallest line
width of the four runs has been selected (since there is no known
mechanism that would artificially {\em de}crease the line width).

{\it Col.\ 8b:} Velocity width in \kms\ of the \HI\ profile measured
at the $25\%$ level of the peak intensity, $\rm\Delta v_{25}$,
corrected for instrumental broadening (\cf Col.\ 6a).

{\it Col.\ 8c:} Velocity width in \kms\ of the \HI\ profile measured
at the $20\%$ level of the peak intensity, $\rm\Delta v_{20}$,
corrected for instrumental broadening (\cf Col.\ 6a).

{\it Col.\ 9a:} Weighted mean \HI\ flux, which is the line integral in
Jy\,\kms , uncorrected for finite beam size.

{\it Col.\ 9b:} The error of the \HI\ flux is the combined error for
the measurement of the profile and the flux calibration: $\sigma
=(\sigma ^2(A) + (0.1 A)^2)^{0.5}$ with $\sigma (A) =
5R^{0.5}A^{0.5}h^{0.5}s^{-1}$ (according to Fouqu\'e \etal \cite{F90})
where $A$ is the line integral and the other parameters as described
for Col.\ 5b.
                
{\it Col.\ 10:} Weighted mean \HI\ peak flux $S_h$ in Jy.

{\it Col.\ 11:} Calculated \HI\ mass in $10^8 d^2_{20}M_{\sun}$ using
$M_{HI} = 2.365 \times 10^5 d^2 S = 948 \times 10^5 S d^2_{20}
M_{\sun}$, where $S$ is the \HI\ flux from Col.\ 9a, and the distance
is $d=20d_{20}$\,Mpc (\cf Sec.~\ref{himass}). No \HI\ masses were
computed for background detections.

{\it Col.\ 12:} The rms noise level $S_{rms}$ over the region used to
fit a baseline, in $10^{-3}$Jy. The smallest values of the four runs
are given here.

{\it Col.\ 13:} The runs when the galaxy was observed.

{\it Col.\ 14:} Notes: the search interval for the possible background
galaxy FCC\,B37 is given. A star denotes interferences in the spectra
which could have prevented us from detecting the line. It is indicated
at the bottom of the table if a velocity is known for that object. A
hash sign indicates if the smallest line width has been discarded
because of either a very low S/N of about three or a large correction
for instrumental broadening ($\ge 7$\kms ). A dagger means that the
diameter has been derived from RC3 values. 

\subsection{Notes on individual objects}

FCC\,2 was classified as a cluster member by Ferguson (\cite{HCF}) but
we have shown it to be a background spiral at $v=4540$\kms .

FCC\,35 shows two unusual-looking peaks. It was confirmed to be at
1800\kms\ by follow-up observations made at Parkes by Staveley-Smith
(1994, priv.\ comm.) and has more recently been observed with the ATCA
(Australia Telescope Compact Array) by Putman \etal (\cite{P98}), who
found that one of the peaks is caused by a nearby intergalactic \HI\
cloud, and the other, as presented here, by the galaxy proper.

FCC\,121 (NGC\,1365) was used as an intra-cluster check of system
performance by separately reducing each 10-minute observation, in each
polarisation band, of this galaxy. Note that only the final averaged
data are given in the table.

FCC\,306 lies in the beam of FCC\,308 and has been confirmed in a
FLAIR observation with $v=915\pm15$\kms\ using emission lines
(Schr\"oder \cite{AS}).

FCC\,338 was classified as S0 in FCC but seems more likely to be an
Sab as given in RC3.

FCC\,B74, the brightest background spiral galaxy in Ferguson's
classifications, has been confirmed with \mbox{$v=4147$\kms }.

\subsection{Comparison with earlier data}

Galaxies in the Fornax cluster have been previously surveyed for \HI\
content by -- among others -- Fouqu\'e \etal (\cite{F90}), Mathewson
\etal (\cite{M92}), Horellou \etal (\cite{H95}), Bureau \etal
(\cite{BMS}), and Barnes \etal (\cite{B97}).  For a detailed listing
of references before 1989 the reader is referred to the HR89
catalogue.

In general, the agreement of our \HI\ radial velocities with published
values is excellent with deviations never exceeding the stated --
usually internal -- errors.  However, \HI\ fluxes vary considerably
with differences more often than not larger than the quoted relative
errors (usually 10\,--\,15\%, \cf\ Sec.~\ref{obs}).  Varying
signal-to-noise ratios for the weaker \HI\ spectra have a significant
impact also on the measured \HI\ line widths.  During our own
observations, we therefore attempted to improve upon the weaker
signals rather than observing a larger number of objects.  Yet, not
all line widths are of the same quality.

We have compared the line widths with published data in the
literature: Bottinelli \etal (\cite{BGFP}), Mathewson \etal
(\cite{M92}), and Bureau \etal (\cite{BMS}) give \HI\ line widths for
Fornax galaxies.  The agreement is within the errors for both 20\% and
50\% line widths except in the case of the 20\% line width of Bureau
\etal which show a mean deviation of $(10\pm3)$\kms\ (for 18 galaxies
in common). The comparison with the other sources indicates that their
line widths are too broad by $\sim 7$\kms\ (after correction for
instrumental broadening).

\section{The \HIb\ content of the Fornax cluster } \label{hi}

In this section we examine the distribution of \HI\ in the Fornax
cluster based on our observations. We first describe the galaxies
detected in \HI\ and then compare their distribution with that of the
galaxies not detected before comparing their \HI\ masses with galaxies
in other environments.

\subsection{The \HIc\ detections } 

\begin{figure}[b]
\vspace{-2.cm}
\resizebox{\hsize}{!}{\includegraphics{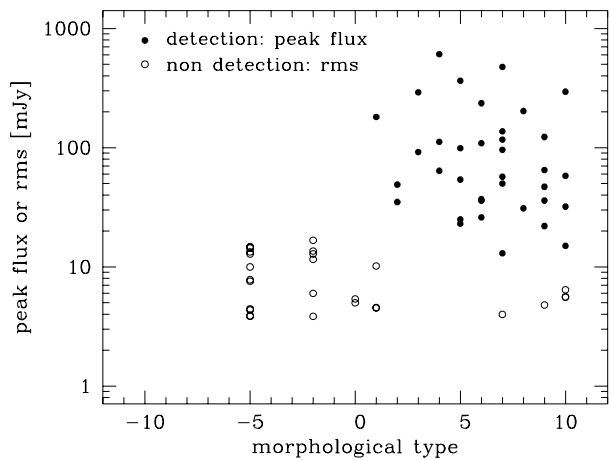}}
\caption[]{Distribution of morphological type for detections (filled
  circles: peak flux is given) and non-detections (open circles), which are
  represented through the rms of the \HIcap\ spectrum.
  }
\label{rmstypplot}
\end{figure}

Our data represent some of the deepest observations ever made of
Fornax cluster galaxies in \HI . We detected 37 of the 66 galaxies, 12
of them for the first time (and two being background
galaxies). Figure~\ref{rmstypplot} shows the morphological types of
our sample (with the numerical coding as in RC3): non-detections are
represented through the standard deviation of the noise (rms) of their
\HI\ spectra (open circles), whereas the peak flux is given for the
detected galaxies. Only a few spiral galaxies were not detected, in
particular three out of four Sa galaxies. Four of the undetected
late-type galaxies are affected by interference at $v\simeq1250$\kms ,
which may have prevented the detection of the galaxy in its vicinity
(no velocities are known from the literature). The fifth galaxy
(FCC\,299) has an optical velocity of ($2051\pm 9$)\kms\ (Schr\"oder
\cite{AS}). A marginal detection at this velocity is possible but
would require at least two more hours of observation to be confirmed.

No elliptical or S0 galaxies ($T\!=\!-2$) were detected. We have
assigned $T\!=\!-5$ to the dwarf spheroidal and Im/dE galaxies since
these classifications indicate early type galaxies.

\subsection{The distribution of \HIc\ detections } 

In the following discussions we have not included the two galaxies
with \HI\ detected at high velocity (FCC~2 and FCC~B74) as they are
not cluster members.

\begin{figure}
\resizebox{\hsize}{!}{\includegraphics{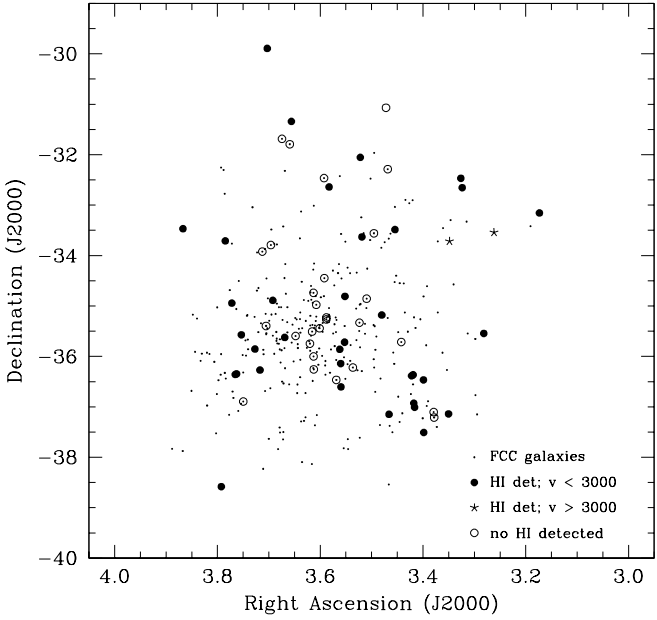}}
\caption[]{The distribution of FCC galaxies (dots), \HIcap\ detections of the
  present work (filled circles) and non-detections (open circles) in J2000
  coordinates. Stars indicate \HIcap\ detections of background galaxies.
  }
\label{maphiplot}
\end{figure}

\begin{figure}
\resizebox{\hsize}{!}{\includegraphics{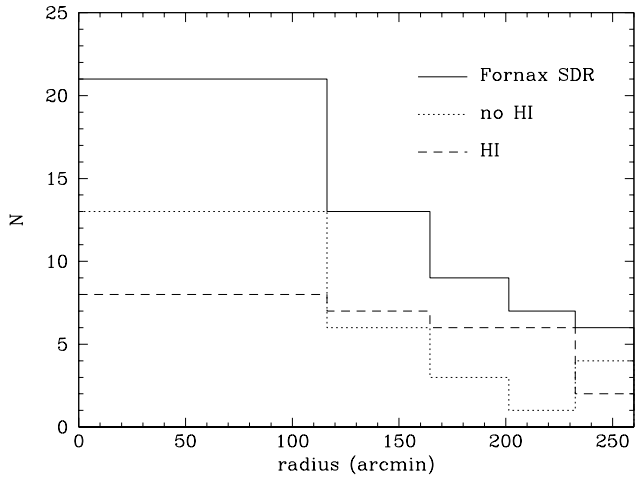}}
\caption[]{The distribution of our total sample, detections (dashed line) and
non-detections (dotted line) in equal area bins as a function of distance from
the cluster centre.
}
\label{radialplot}
\end{figure}

The present paper gives the most comprehensive list of \HI\ detections
in the Fornax cluster to date. In Fig.~\ref{maphiplot} we show the
distribution of FCC galaxies in the sky, with the \HI\ detections as
filled circles and the non-detections as open circles. Note that our
sample includes four galaxies outside the FCC as well as three
background galaxies of which two are not in the FCC main catalogue. To
quantify we plot in Fig.~\ref{radialplot} the numbers of galaxies in
radial annuli of equal area for the total sample and the galaxies with
and without \HI\ detections.

\begin{figure}
\resizebox{\hsize}{!}{\includegraphics{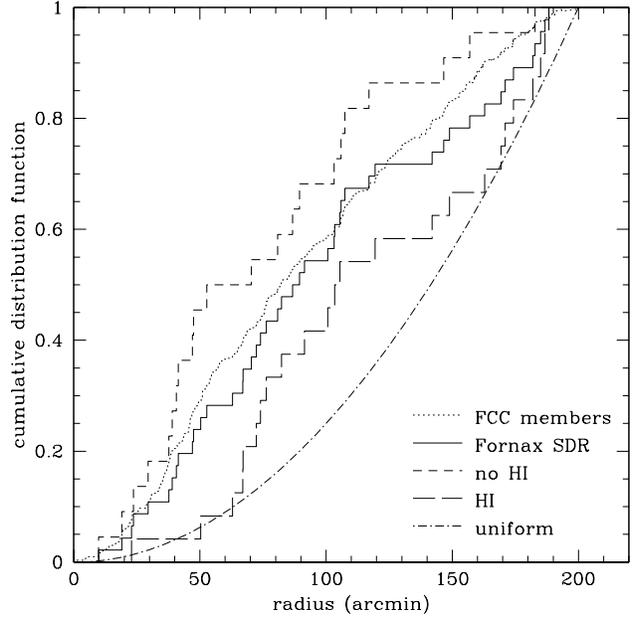}}
\caption[]{Cumulative radial distribution of our total sample (solid
  line) compared to the distribution of all cluster members in the optical FCC
  sample (dotted line). The subsamples of our galaxies with and without any
  detected \HIcap\ are indicated by long and short dashed lines
  respectively. A uniform, constant-density distribution is also shown for
  comparison.
  }
\label{cdfplot}
\end{figure}

Both Figures~\ref{maphiplot} and \ref{radialplot} show that the
galaxies with \HI\ detections are evenly distributed across the
cluster, whereas those without any detected \HI\ are much more
concentrated towards the cluster centre. This is to a large extent a
reflection of the usual density-morphology relation: the ratio of
early to late-type galaxies increases rapidly towards the cluster
core. In Fig.~\ref{radialplot} the distribution of galaxies with
detected \HI\ is consistent with a constant surface density as though
these galaxies were not aware of the cluster, although this is partly
due to the binning: see the cumulative distribution in
Fig.~\ref{cdfplot}.

In Fig.~\ref{cdfplot} we plot the same data as cumulative distribution
functions, also including cluster members listed in the optical FCC
catalogue. We can now assess the significance of the difference in the
distributions. The radial distribution of our total sample is not
significantly different from that of the FCC members: they differ at a
Kolmogorov-Smirnov (KS) significance of only 15\%. There is a
significant difference in the radial distributions of the \HI\ and
non-\HI\ samples: they differ at a KS significance of 97\% with the
non-detected galaxies being much more concentrated towards the cluster
centre. Interestingly, the distribution of the \HI\ detected galaxies
is also more centrally concentrated than a uniform distribution
(dash-dotted curve in Fig.~\ref{cdfplot}) at a KS significance of
94\%. The \HI -rich galaxies are not distributed as a uniform sheet
but clearly show signs of being concentrated towards the cluster
although at larger radii than the non-detected galaxies.

We also compared the velocity distribution of the cluster galaxies
detected in \HI\ with published velocities of the other cluster
members in our sample. The 35 cluster galaxies detected in \HI\ have a
mean velocity of $1480\pm 60$\kms\ and the 26 members not detected in
\HI\ with published velocities have a mean velocity of $1460\pm
70$\kms. There is no significant difference in these velocity
distributions.

\subsection{The \HIc\ masses } \label{himass}

Estimates of the \HI\ mass depend on the cluster distance, which is
currently under discussion. There are Cepheid distances available for
three Fornax galaxies: NGC\,1365 at 18.6\,Mpc (Madore \etal
\cite{mad1999}), NGC\,1326A at 18.7\,Mpc (Prosser \etal
\cite{pro1999}), and NGC\,1425 at 22.2\,Mpc (Mould \etal
\cite{mou2000}). Drinkwater \etal (\cite{mjd}) have shown that
NGC\,1326A is part of a subcluster in front of the main cluster (see
Section~\ref{tf}) and they also argue that NGC\,1365 is in front of
the cluster core. They therefore propose -- as did Mould \etal
(\cite{mou2000}) -- that the mean of the three distances, 20\,Mpc, be
adopted for the Fornax cluster. We note that there is still no
consensus in the literature on this matter, and for this paper we
parameterise the distance as $d=20d_{20}$\,Mpc.

The \HI\ masses for our galaxies are calculated using $M_{HI} = 2.37
\times 10^5 d^2 S = 948 \times 10^5 S d^2_{20} M_{\sun}$, where $S$ is
the \HI\ flux integral in Jy\kms. These distance-parameterised masses
are listed in Table~2 in units of $d_{20}^{2} M_{\sun}$.

\begin{figure}
\resizebox{\hsize}{!}{\includegraphics{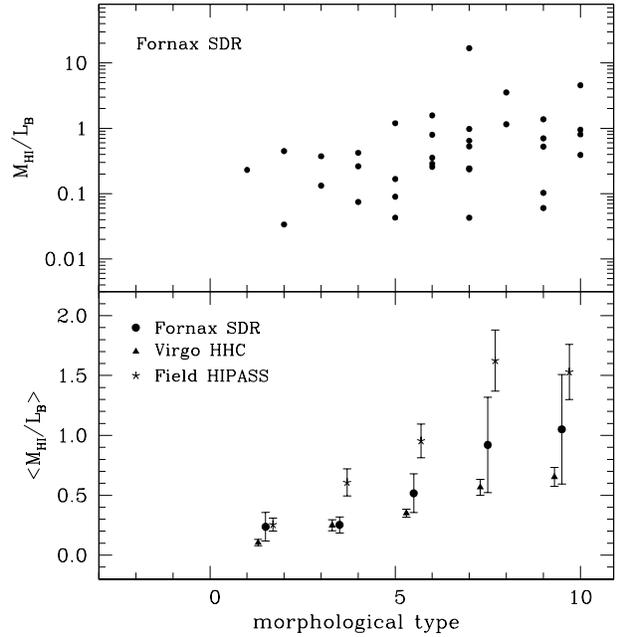}}
\caption[]{The \HIcap\ mass-to-blue light ratio versus morphological
  type for our sample. The upper panel shows the distribution of the
  sample in logarithmic scale, the lower panel shows the binned data in
  comparison with the Virgo cluster (triangles) and with the field
  (stars). For clarification the points are slightly offset in types.
  }
\label{mhilbplot}
\end{figure}

NGC\,1365 ($T\!=\!4$) has the largest \HI\ mass in Fornax and is one
of the most \HI -massive galaxies known. The second most \HI -massive
galaxy in Fornax, NGC\,1326B ($T\!=\!7$), is one member of an
interacting pair. Its \HI\ mass-to-light ratio is extremely large,
presumably due to its interactions (\cf Fig.~\ref{mhilbplot}). Its
companion, NGC\,1326A ($T\!=\!5$), shows no unusual parameters.

NGC\,1316C has the lowest \HI\ mass for the Sd galaxies. It is a close
neighbour of FCC\,35 ($T\!=\!9$) which is associated with an \HI\
cloud (Putman \etal \cite{P98}). This \HI\ cloud may have been
stripped from NGC\,1316C by a close encounter with FCC\,35, which has
an \HI\ mass consistent with its morphological type.


The distance-independent \HI\ mass-to-blue light ratio \ml\ is shown
in the upper panel of Fig.~\ref{mhilbplot} in logarithmic scale. We
have binned the data in pairs of morphological types for better
statistics (1--2, 3--4, 5--6, 9--10) and calculated the mean. The
result is shown in the lower panel of Fig.~\ref{mhilbplot}. NGC\,1326B
with its extreme \ml\ has been excluded from this mean.

In comparison we show the means of galaxies in the Virgo cluster
(triangles), as derived from data obtained with the Arecibo telescope
(Helou \etal \cite{HHC1}; Helou \etal \cite{HHC2}; Hoffman \etal
\cite{ HHC3}; Hoffman \etal \cite{HHC4}; for simplicity we shall call
this the HHC data, as in Helou, Hoffman and collaborators). For the
calculation of the mean we have again excluded an unusually large \ml\
(for $T\!=\!3$) in this data-set. Despite the larger errors for the
Fornax sample (due to small sample statistics) we can see that the
Virgo galaxies, well-known for their \HI\ depletion, have
systematically lower \ml\ ratios. The mean \ml\ for the Fornax sample
is $0.68\pm0.15$ whereas that of the Virgo sample is
$0.45\pm0.03$. The means differ at the 85\% confidence level according
to the T-test.

Finally, we also show the means from data obtained in the \HI\ Parkes
All-Sky Survey (HIPASS) of bright field galaxies in the southern
hemisphere (Marquarding \cite{MM00}; see also Staveley-Smith \etal
\cite{SS96} for a description of the survey). The \HI\ selected field
galaxies (indicated with stars) show systematically larger \ml\ ratios
than the galaxies in either cluster. The mean \ml\ for the field
sample is $1.15\pm0.10$ which is larger than the Fornax value at the
98\% confidence level and larger than the Virgo value at a confidence
level greater than 99.9\% (both according to the T-test).

Galactic extinction corrections have been applied to the HIPASS
sample; in case of Fornax and Virgo we assume the Galactic extinction
to be zero or negligible. Since the HIPASS sample does not have
inclinations for all galaxies we have not corrected the blue
luminosities for any of the samples for internal absorption. A
correction for internal absorption will lower the mean value of \ml .
The HIPASS sample is limited by peak flux and therefore biased against
largely inclined galaxies. However, the correction in this case would
be smaller than for the cluster samples with their more evenly
distributed inclinations and the difference between field and cluster
galaxies will be increased.


\subsection{\HIc\ deficiency }

Solanes \etal (\cite{Sol96}, hereafter SGH) have shown that the \HI\
content of isolated spiral galaxies not only depends on morphological
type but also shows a tight linear relation with the linear optical
diameter. We can therefore compare the actual observed \HI\ mass as
derived from our \HI\ fluxes with the expected value for a galaxy
unaffected by environmental conditions and of the same diameter and
morphological type. Using the tabulated coefficients for the
expectation value of the logarithm of the \HI\ mass, $\log
\widehat{M}(T,D)$, from SGH, we have calculated the \HI\ deficiencies
for our galaxies as
\[ {\rm HI}_{def} = \log {\widehat{M}}_{HI}(T^{obs},D^{obs}_{opt}) - \log M^{obs}_{HI}, \]
where the diameters are given in kpc and the \HI\ masses in solar
units. We have used the diameters from the FCC (measured at the 26.5th
isophote), and corrected RC3 diameters for 3 galaxies outside of the
FCC by adding the mean difference of 0.05. Since SGH have used
UGC\footnote{Uppsala General Catalog of Galaxies (Nilson \cite{ugc}).}
diameters, we have then adjusted the conversion given by Horellou
\etal (\cite{H95}) to compute UGC from our FCC diameters:
\[ \log (D_{UGC}+0.3) = 1.0173 \log D_{25}. \]

Because of the additional correction for diameter dependence, we would
expect the \HI\ deficiencies to have a smaller internal scatter and be
a better indicator for variations in \HI\ content than the \ml\
values. However, in addition to the (sometimes large) errors in
observed \HI\ flux, diameter and morphological type, both the \HI\
mass and the linear diameters are distance dependent, which increases
the scatter in the \HI\ deficiencies again. While the same distance
can be used for a small nearby cluster as Fornax, the distances to the
field galaxies used for the expectation value by SGH depend on the
Hubble constant and show errors due to peculiar velocities. Solanes
\etal (\cite{Sol}) therefore emphasise that \HI\ deficiencies are only
meaningful in a statistical sense.

Expectation values of \HI\ masses are only well determined for types
\mbox{$T\!=\!1$\,--\,5}. As indicated by Solanes \etal ({\cite{Sol}),
we have calculated the expectation values of \HI\ masses for the later
types ($T\!=\!6$\,--\,10) using the coefficients for $T\!=\!5$. Though
the variation in the coefficients with morphological type are small
(\cf Haynes \& Giovanelli \cite{HG84}), we want to emphasise that \HI\
deficiencies for types (much) later than 6 are less reliable.

Figure~\ref{hidefplot} shows the \HI\ deficiencies of all our Fornax
galaxies with \HI\ measurements. The scatter is large and the number
of galaxies for each morphological type is small. We have therefore
derived a mean \HI\ deficiency for the types $T\!=\!1$\,--\,5 together:
$<\!{\rm HI}_{def}\!>=0.35 \pm 0.13$. Including galaxies with $T\!=\!6$,
we find the similar value of $<\!{\rm HI}_{def}\!>=0.38 \pm 0.09$. If
we assume that the deficiency is reliably defined for the later types
as well, we find $<\!{\rm HI}_{def}\!>=0.30 \pm 0.07$, which is close
to the other values. These results show a significant if moderate
\HI\ deficiency for the Fornax cluster, independent of our \ml\
calculations above. Our deficiency value is consistent with that of
Horellou \etal (\cite{H95}) who derived a deficiency of $0.43\pm
0.55$, with a much larger uncertainty.

\begin{figure}
\vspace{-2.3cm}
\resizebox{\hsize}{!}{\includegraphics{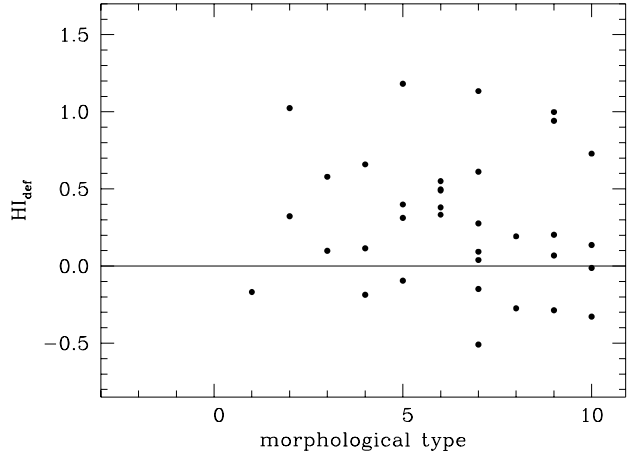}}
\caption[]{The \HIcap\ deficiency parameter versus morphological type
  for our sample. The expected \HIcap\ masses for galaxies with $T\ge 6$ were
  calculated using the relationship for galaxies with $T\!=\!5$.
  }
\label{hidefplot}
\end{figure}

We have calculated \HI\ deficiencies for the HIPASS sample as well,
using RC3 diameters where available (for 106 out of 136 galaxies). The
mean deficiencies for all types together as well as for each type
separately are consistent with zero (\eg $0.05 \pm 0.03$ for
$T\!=\!1$\,--\,6), indicating that the sample, though \HI\ selected
and not optically selected, is representative for the field. The
scatter for $T\!=\!1$\,--\,6 is 0.24 and comparable to the scatter of
0.23 quoted by SGH for their field galaxies.

\subsection{Discussion}

Environmental effects on the \HI\ content of spiral galaxies have been
discussed at length in the literature (\eg Solanes \etal \cite{Sol};
Abadi \etal \cite{A99}; van Gorkom \cite{vG96}; Giovanelli \& Haynes
\cite{GH85}; and many others).  There are two main mechanisms that can
strip galaxies of their gas content: ram-pressure stripping by
interaction with hot intra-cluster gas, and tidal interactions with
other galaxies. The second mechanism can also strip stars off the
outer parts of a galaxy. Other mechanisms are viscous stripping and
thermal conduction.  Rich galaxy clusters, like Virgo (Solanes \etal
\cite{Sol}; Cayatte \etal \cite{C94}; Guhathakurta \etal \cite{G88};
Hoffman \etal \cite{HHS}; to name only a few) or Coma (Bravo-Alfaro
\etal \cite{BA00}; Bothun \etal \cite{B84}) have large X-ray
luminosities, and galaxies in the inner part of the clusters show
definite signs of loss of \HI\ gas due to ram-pressure stripping.

Fornax is a small cluster but with a high central galaxy density. Its
X-ray luminosity is much lower than that of the Virgo cluster (about 2
orders of magnitude) with a significant detection only in the core of
the cluster (Killeen \& Bicknell \cite{kb88}; Jones \etal
\cite{J97}). Within the core region however (a projected radius of
about 20 arcmin) the mean gas density is of order $10^{-3}$cm$^{-3}$
(Jones et al.) so pressure stripping will be effective with timescales
of $10^9$ years according to the conservative estimates of Ferguson \&
Binggeli (\cite{FB94}). There is also direct evidence for stripping in
some of the central Fornax galaxies from their morphologies, e.g.\
NGC~1404 (Jones \etal \cite{J97}). The effect of pressure stripping is
harder to calculate at larger radii, but it is presumably much
lower. In a large sample of galaxies from different clusters (most
well-outside the core region), Solanes \etal (\cite{Sol}) found no
apparent relationship between (strong) X-ray luminosity and \HI\
deficiency in individual galaxies. (However the fraction of gaseous
late-type spirals in the centre of rich clusters may be reduced in
favour of lenticulars that have no or only little gas content.)

Our Fornax data demonstrate a non-zero \HI\ deficiency parameter as
well as a mean \ml\ ratio significantly lower than that found for a
sample of field galaxies. This is for a sample of galaxies that
extends well beyond the core where we might expect pressure stripping
to be important. This result appears to conflict with previous
non-detections of any {\em strong} \HI\ deficiency in Fornax (Horellou
\etal \cite{H95}, and Bureau \etal \cite{BMS}). In the former case the
sample actually measured was very small (only 6 new measurements of
cluster galaxies), which resulted in a large uncertainty, and they
were mostly earlier types where the depletion is weaker. In the case
of the Bureau \etal result the conclusion was based on a comparison
with galaxies in the Ursa Major cluster. We note that the Fornax
cluster is not significantly denser than Ursa Major if the central
core region is excluded and that we might therefore expect them to
have similar \HI\ deficiencies. Our use of a large sample and our
comparison with a genuine field sample from the HIPASS data
(Marquarding \cite{MM00}) has allowed us to detect the \HI\ deficiency
in the Fornax cluster.

As we describe above, a number of mechanisms have been proposed that
could explain the \HI\ deficiency of Fornax cluster galaxies. Most of
the galaxies we detected in \HI\ are well beyond the cluster core
where the pressure of the hot X-ray gas is high enough for
ram-pressure stripping. However, as discussed by Solanes \etal
(\cite{Sol}), this mechanism is still viable if the galaxies are on
radial orbits in the cluster. Our sample is too small for a detailed
analysis of the orbits, but we can make a simple comparison of the
kinematics of the \HI-deficient galaxies (${\rm HI}_{def}>0.3$,
n\,=\,17) with the rest of the \HI\ detected galaxies. The \HI
-deficient galaxies have a smaller velocity dispersion ($247\pm44$\kms
) than the other galaxies ($391\pm67$\kms ) with the difference
significant at the 93\% confidence level. Splitting the sample
according to \ml\ ratio gives very similar results. This difference in
velocity dispersions is consistent with the deficient galaxies having
more radial orbits than the other galaxies as was shown for a
composite cluster sample by Solanes \etal (\cite{Sol}). One further
process that will contribute to \HI\ depletion is the conversion of
neutral gas into stars.  Our optical measurements (Drinkwater \etal
\cite{mjd01b}) have shown that there is an excess of star formation in
Fornax cluster galaxies at similar distances to most of our \HI\
detections.  Although the corresponding gas depletion timescales are
long (of order $10^{10}$ years), this does show that the cluster
environment can influence the gas content of galaxies even at these
large distances.

\section{The \tfr\ } \label{tf}

During our observing programme we observed all spiral galaxies in the
FCC with long integration times. In combination with the multi-colour
multi-aperture data obtained by one of us (Schr\"oder \& Visvanathan
\cite{SV}; Schr\"oder \cite{AS}) this comprehensive sample enables us
to establish the \tfr\ for all inclined spiral galaxies ($i \ge
45\degr$, $1\le T \le 9$) in the Fornax cluster. We have taken the
parameters as well as the corrections as described in Schr\"oder
(\cite{AS}): morphological types are from the FCC, axial ratios are
mainly from Lauberts \& Valentijn (\cite{LV}, \cf notes to Table~2),
$q_0$ to calculate inclinations are from Heidmann \etal (\cite{Heid}),
and internal absorptions have been corrected according to the RC3. We
have taken the 20\% line widths from Tab.~2 (which are corrected for
instrumental broadening), and applied a correction for z-motion and
turbulences of $-12.6$\kms\ as explained in Richter \& Huchtmeier
(\cite{RH}). For the \tfr\ we use the maximum rotational velocity $V_M
= \Delta v_{20}^c (2 \sin i)^{-1}$.  Adopting a slope of $-6.50$ from
the Virgo cluster (Schr\"oder \cite{AS}) where we have better
statistics, we find
\[ B_T^{o,i}\,(F) = -6.50 \cdot [\log V_M-2.0] + (12.56 \pm 0.18) \] 
with a scatter of 0.84 mag. 

Excluding the faintest galaxy from the relation, FCC\,306 (\cf
Fig.~\ref{tfforplot}), we find a scatter of 0.71, which is
considerably smaller. No obvious explanation for such a deviation
could be found: This Sm galaxy has a strong \HI\ signal, and the
velocity has been confirmed with an optical measurement (Schr\"oder
\cite{AS}). Its 20\%-width is possibly smaller by $\sim\!15$\kms\
(Waugh, priv. comm., using HIPASS data) which would result in a
scatter of 0.81. Its \HI\ mass and \ml\ is consistent with other
galaxies of this type. The optical data is good (four apertures
covering a large range in radius of the galaxy) and the total
magnitude is comparable to the magnitude given by Lauberts \&
Valentijn (\cite{LV}). The axial ratio is confirmed by Loveday
(\cite{L96}). However, we believe that its unusually large effect on
the scatter of the \tfr\ justifies sufficiently its exclusion from
here on from the relation. We assume that a combination of errors in
the various parameters accounts for its unusual high deviation from
the \tfr .  However, another explanation would be that this galaxy
lies off the main \tfr\ because it actually lies behind the main
cluster and and is falling into it. This argument is supported by the
very compact morphology of FCC~306: it has the smallest angular size
and highest surface brightness of the star-forming dwarfs in the
sample of Drinkwater \etal (\cite{mjd01b}).

Due to its small velocity dispersion the Fornax cluster has so far
been assumed to have no substructure. However, Drinkwater \etal
(\cite{mjd}) have analysed the dynamics of the cluster and find
definite evidence for a subcluster in the South West of the cluster,
centred on Fornax~A (NGC\,1316). In fact, a slightly enhanced galaxy
density in this region can also be seen in Fig.~16 of the FCC.  If we
exclude the galaxies in this subcluster from the \tfr\ we obtain
\[ B_T^{o,i}\,(F) = -6.50 \cdot [\log V_M-2.0] + (12.69 \pm 0.12), \] 
as shown in Fig.~\ref{tfforplot}, with a scatter of 0.50. The scatter
is considerably smaller than before and comparable to the intrinsic
scatter, which is about 0.4 (\cf Sakai \etal \cite{S00} find 0.43 for
near-by galaxies with Cepheid distances).

\begin{figure}
\vspace{-2.3cm}
\resizebox{\hsize}{!}{\includegraphics{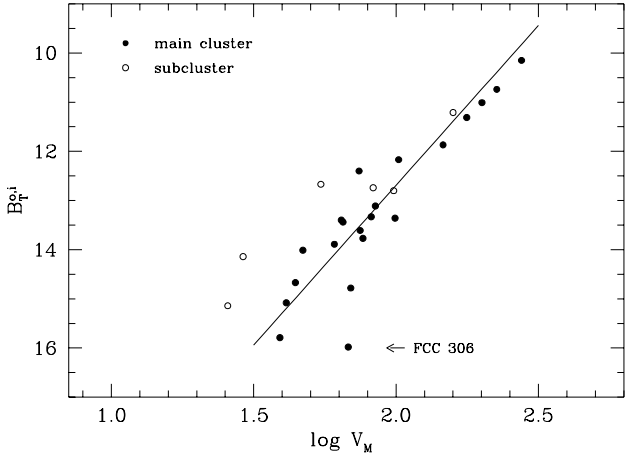}}
\caption[]{The \tfr\ for the Fornax cluster. Open circles denote galaxies
  which seem to belong to a subcluster. The linear regression is derived for
  the main cluster only and without FCC\,306.
  }
\label{tfforplot}
\end{figure}

We have not attempted to derive a \tfr\ of the subcluster alone since
only six galaxies therein are useful for the \tfr . Furthermore, one
of these has unusual \HI\ masses due to interactions (NGC\,1316C, \cf
Sec.~\ref{hi}). However, in the mean they are $(1.0 \pm0.5)$\,mag
brighter than the main cluster, indicating that the subcluster lies in
the foreground of Fornax (which is also supported by the lower Cepheid
distance of NGC\,1326A, 18.7\,Mpc, in the subcluster). In combination
with its slightly larger mean velocity of 1583\kms\ (as compared to
1478\kms\ for the main cluster, Drinkwater \etal \cite{mjd}) we now
have a three-dimensional picture of the subcluster in the foreground
falling into the main cluster. The high rate of \HI\ detection in the
subcluster (see Fig.~\ref{maphiplot}) suggests that it is falling into
the main cluster for the first time. This is consistent with the
slightly lower \HI\ deficiency ($0.2\pm0.2$) we measure for the
subcluster galaxies compared to the rest of the cluster
($0.44\pm0.10$).

The \tfr\ for the main cluster can be used to derive a relative
distance to the Virgo cluster as shown by Schr\"oder
(\cite{AS}). Using the same parameters and corrections where possible
(see description above; morphological types and axial ratios were
taken from the VCC\footnote{Virgo Cluster Catalog (Binggeli \etal
\cite{vcc}).}, \HI\ line widths from Bottinelli \etal \cite{BGFP}) we
find
\[ B_T^{o,i}\,(V) = -(6.50\pm0.36) \cdot [\log V_M-2.0] + (13.07 \pm 0.07) \] 
for Virgo, with a scatter of 0.64 mag. This relation includes all
galaxies classified as members of the Virgo cluster proper (Binggeli
\etal \cite{B93}). We find a relative distance of $-0.38\pm 0.14$\,mag
with the Fornax cluster being closer. This is consistent with other
\tfd s between Fornax and Virgo (Bureau \etal (\cite{BMS}):
$-0.06\pm0.15$; Aaronson \etal (\cite{A89}): $-0.25\pm0.23$;
Visvanathan (\cite{v83}): $-0.20\pm0.18$), contrary to some other
distance measurements that place Fornax further away (\eg McMillan
\etal (\cite{M93}): $0.24\pm0.10$, using the planetary nebulae
luminosity function).  The relative distance measurements found in the
literature vary from $-0.5$\,mag to $+0.4$\,mag (see Tab.\ 6.1 in
Schr\"oder \cite{AS}; Tab.\ 3 in Bureau \etal \cite{BMS}). However,
many methods use only a small sample of galaxies (\eg type Ia
supernovae, surface brightness fluctuations, planetary nebulae
luminosity functions) where the cluster centres are less well defined.

The \tfr\ of the spatially more confined Virgo subcluster B with 20
galaxies shows a smaller scatter of 0.46, similar to the one of the
Fornax cluster galaxies and of the calibrators. This agrees well with
the velocity dispersion of the B cluster being significantly smaller
than that of the whole Virgo cluster (499\kms\ versus 699\kms, see
Binggeli \etal \cite{B93}). We find that the local intercept is
fainter ($13.21\pm0.11$) and the relative distance larger: $-0.52\pm
0.16$\,mag.

\section{Conclusion}

In this paper we have presented deep \HI\ observations of all spiral
galaxies as well as bright early-type galaxies in the FCC. Two
late-type spiral galaxies and three irregulars were not detected,
possibly because of an interference at $v\simeq1250$\kms . Only one
out of four Sa galaxies was detected, and none of the S0/a galaxies
and galaxies of earlier morphological types.

The distribution of galaxies with \HI\ in the Fornax cluster differs
significantly from the more centrally concentrated distribution of
non-\HI\ detections, as expected from the density-morphology relation.
Even the \HI -rich galaxies are more centrally concentrated around the
cluster than a random distribution: this shows that they are aware of
the cluster potential. However, there is no significant difference in
the velocity distribution of the two samples.

The mean \ml\ binned by morphological type of the Fornax galaxies is
between those of the Virgo galaxies and of galaxies in the field
indicating a modest but significant \HI\ depletion: the mean \ml\ is
$60\pm 13$\% the value for a comparison sample of field galaxies. In
addition, the \HI\ deficiency parameter, as introduced by Solanes
\etal (\cite{Sol96}), is $0.38\pm0.09$, which is significantly greater
than zero.  There is some indication from the kinematics of the
galaxies that this \HI\ depletion is caused by ram-pressure stripping
of galaxies which are on orbits that pass closer to the cluster
core. In addition, optical observations show evidence of enhanced star
formation in galaxies in these outer parts of the cluster which will
use up some of the gas.

We have calculated the \B -band \tfr\ relation for the Fornax cluster
from our data and obtain a good fit for the main cluster with a
scatter of 0.50\,mag. The relative distance to the Virgo cluster is
$-0.38\pm 0.14$\,mag with the Fornax cluster being closer.  The \tfr\
confirms the existence of a subcluster in the South-West of the main
cluster, centred on NGC\,1316 (Fornax A).  The subcluster galaxies are
almost one magnitude brighter than the main cluster. Combined with
their higher mean velocity and the higher \HI\ detection rate, this
indicates that the subcluster lies in the foreground and is currently
falling into the cluster for the first time.

\bigskip

\acknowledgements 

We thank the staff at the Parkes Observatory, in particular Euan
Troop, for making our stays at the telescope very pleasant and
productive.  Several people helped us in conducting the observations
at the Parkes telescope, notably M.G.\ McMaster (STScI), R.\ Otrupcek
(ATNF), R.C.\ Kraan-Korteweg (Univ.\ of Guanajuato), and P.\ Henning
(Univ.\ of New Mexico). We are very grateful to M.\ Waugh and M.\
Marquarding for making some HIPASS measurements available prior to
publication, and to M.\ Zwaan for helpful discussions. We wish to
thank the referee of this paper for detailed comments which greatly
improved the presentation of this work.  A. S. gratefully acknowledges
financial support from the European Marie Curie Fellowship Grant.


\vfill


\appendix

\section{\HI\ spectra}

The \HI\ spectra (uncalibrated) for all galaxies detected during our
Parkes observations are shown in Fig.~\ref{fccplot}. The calibration factor varied
from year to year: 1.00 for 1991, 1.25 for 1992, 1.28 for 1993, and
1.32 for 1994. We show one spectrum per galaxy (because of the
different set-ups per year we did not average the spectra of the
different years), the others are available from one of the authors
(Schr\"oder).

During the 1993 and 1994 observing runs, data suffered frequently from
strong interference at a central radial velocity of 1250\kms .
Negative spikes were excluded by range selection but positive spikes
are sometimes easily visible in the plotted spectra.

\begin{figure*}
\resizebox{\hsize}{!}{\includegraphics{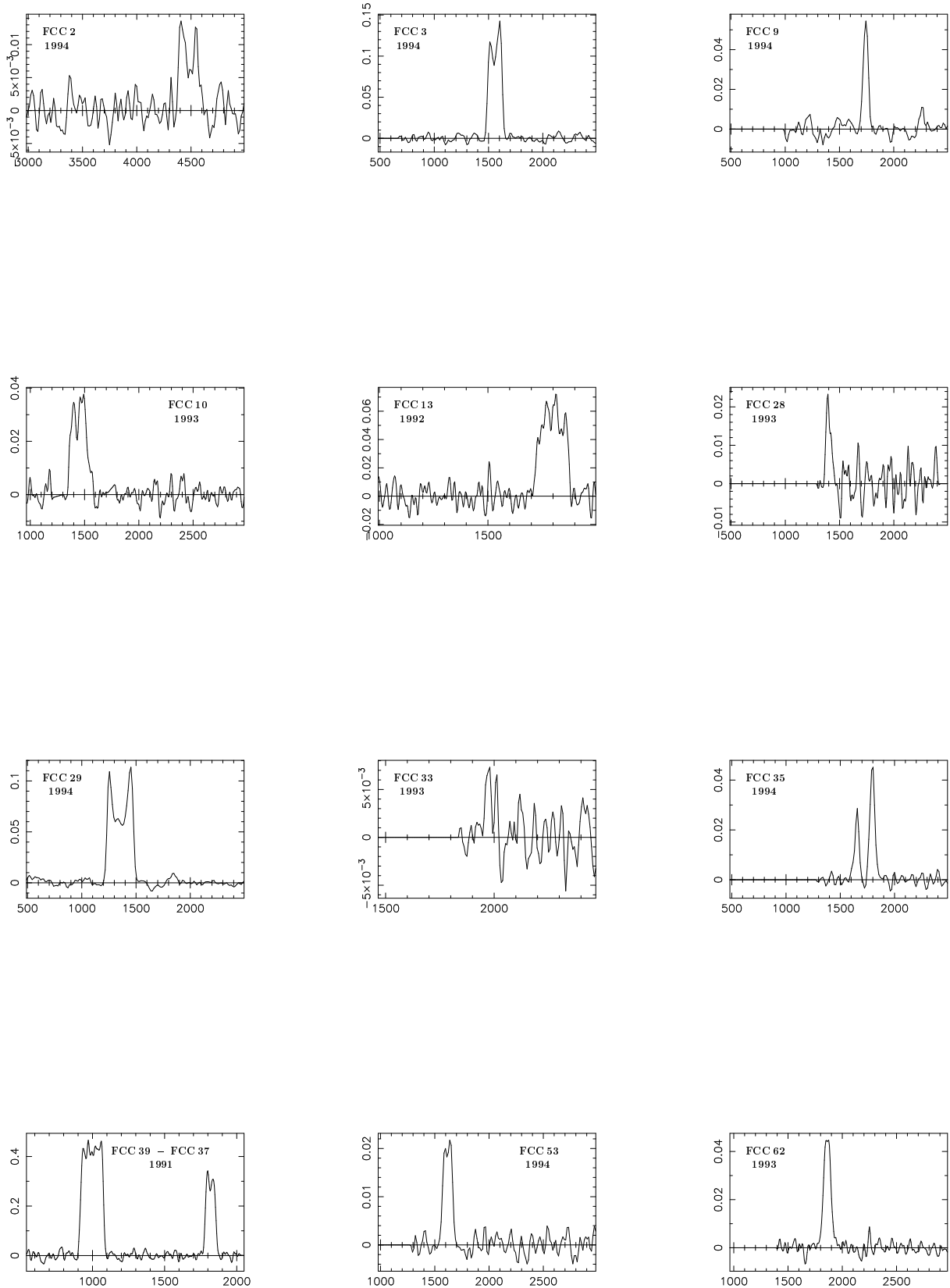}} \vspace{-2.5cm}
\caption[]{\ion{H}{i} spectra of the detected galaxies in the
  Fornax cluster. In each plot the flux in mJy is plotted versus the
  velocity (optical convention) in km\,s$^{-1}$.  
 }
\label{fccplot}
\end{figure*}

\addtocounter{figure}{-1}
\begin{figure*}
\resizebox{\hsize}{!}{\includegraphics{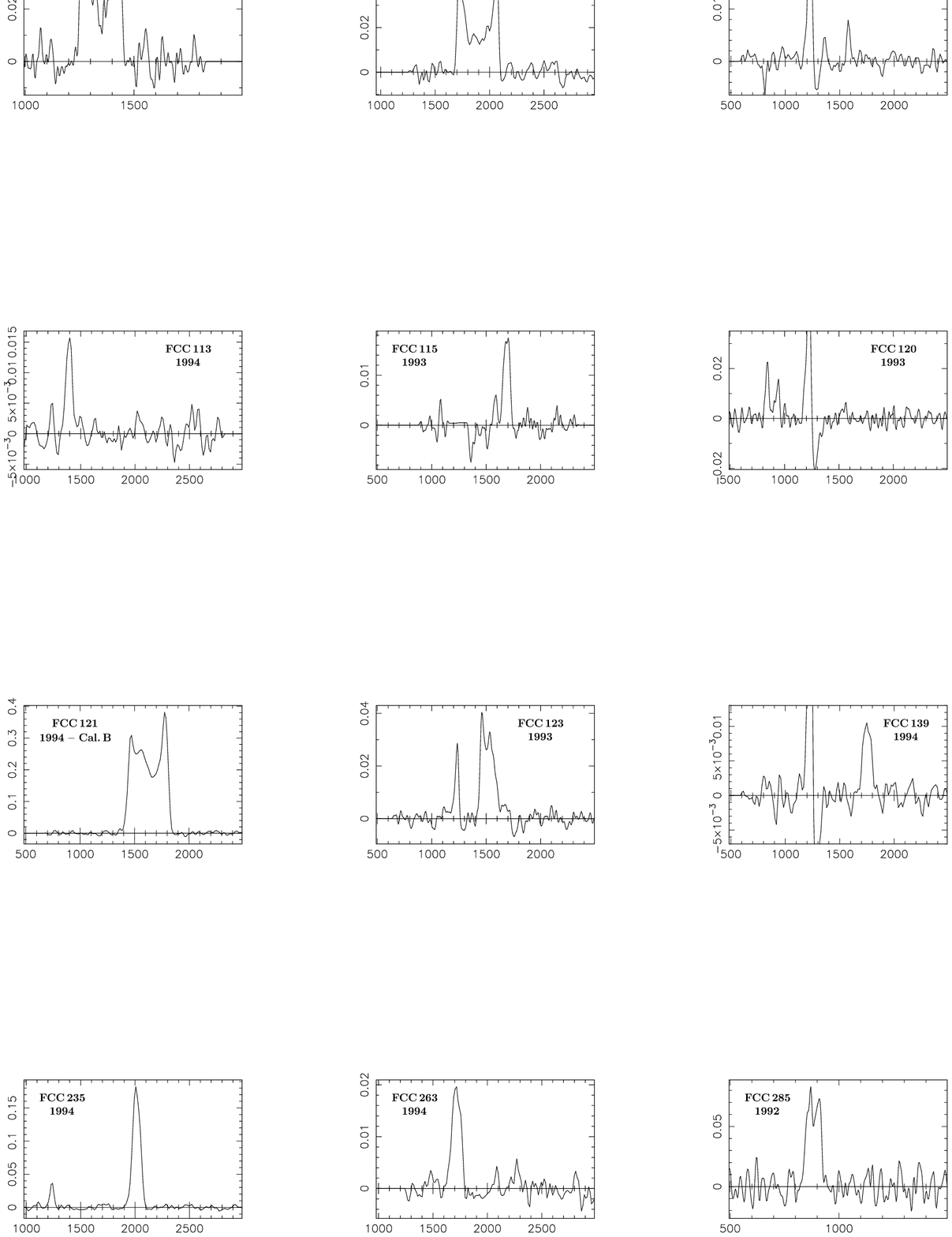}} \vspace{-2.5cm}
\caption[]{continued }
\end{figure*}

\addtocounter{figure}{-1}
\begin{figure*}
\resizebox{\hsize}{!}{\includegraphics{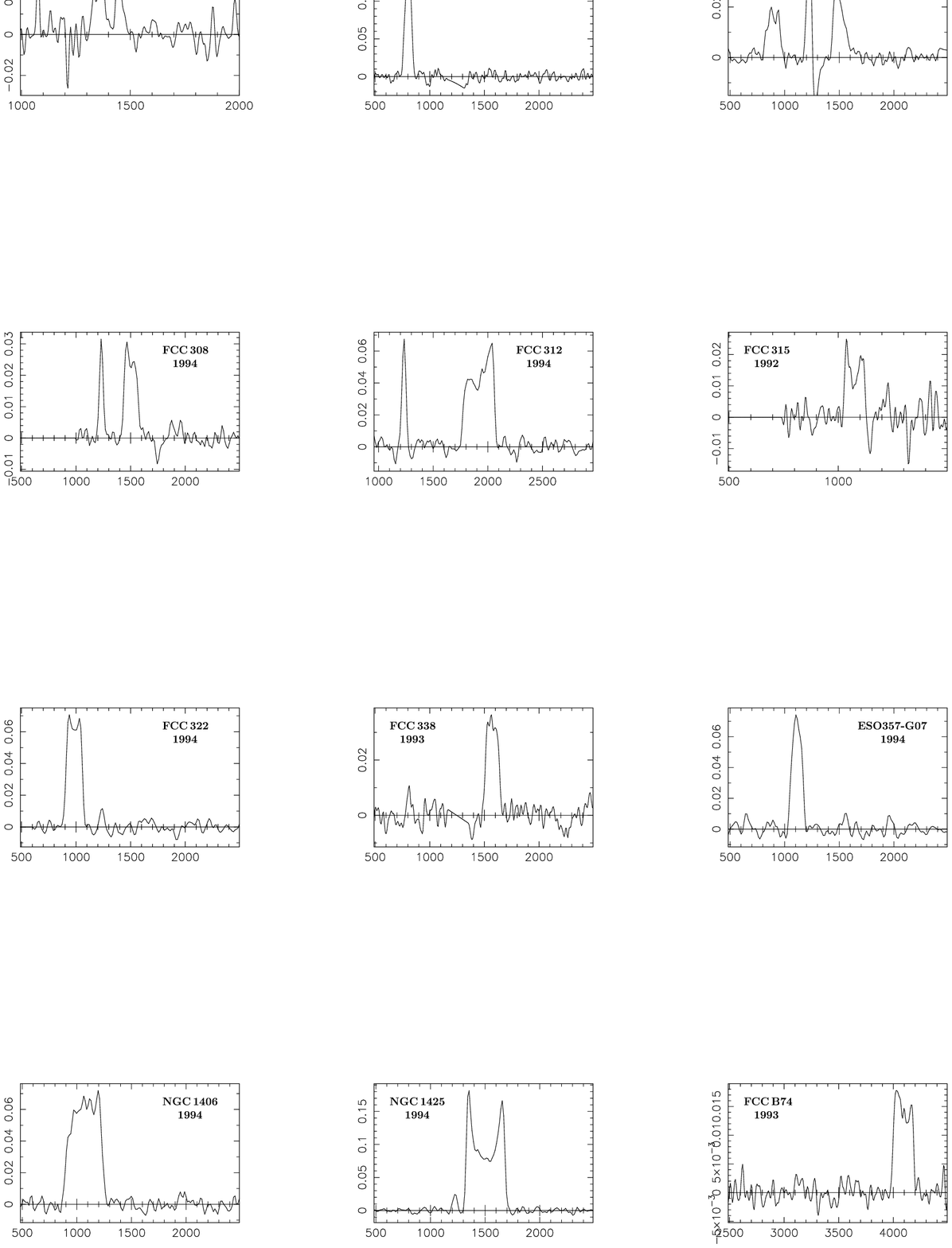}} \vspace{-2.5cm}
\caption[]{continued }
\end{figure*}



\end{document}